
\input harvmac
\def\labelfile{labeldefs.tmp}
\immediate\openin\lfile\labelfile\relax
\ifeof\lfile\immediate\closein\lfile
\else\expandafter\input\expandafter\labelfile\fi\writedefs
\def\future#1{\ifx#1\undfined\message{\noexpand#1 must be defined later.}\else
\expandafter#1\fi}
\Title{LA-UR-93-1153}{DEEP INELASTIC ELECTROPION
PRODUCTION*}\footnote{}{*Research supported by NATO Science \& Environmental
Affairs Division and \vskip0pt\hskip2pt{The DOE High Energy Physics Division}}


\centerline{A. Calogeracos}
\centerline{School of Engineering}
\centerline{University of Thessaly}\centerline{Volos 383 34, Greece}
\bigskip\centerline{Norman Dombey\footnote{$^\dagger$}
{(normand@syma.sussex.ac.uk)}}
\centerline{Physics Division}
\centerline{University of Sussex}\centerline{Brighton,
Sussex}\centerline{United Kingdom BN1 9QH}
\bigskip\centerline{Geoffrey B.
West\footnote{$^\ddagger$}{(gbw@pion.lanl.gov)}}
\centerline{High Energy Physics, T-8, MS B285}
\centerline{Los Alamos National Laboratory}\centerline{Los Alamos, NM
87545}\centerline{U. S. A.}


\vfill\eject
\centerline{\bf ABSTRACT}
\smallskip
{This paper is devoted to a study of possible scaling laws, and their
logarithmic corrections, occurring in deep inelastic electropion production.
Both the
exclusive and semi-exclusive processes are considered.  Scaling laws,
originally motivated from PCAC and current algebra considerations are examined,
first
in the framework of the parton model and QCD peturbation theory and then from
the more formal perspective of the operator product expansion and asymptotic
freedom, (as expressed through the renormalization group).  We emphasize that
these processes allow scaling to be probed for the full amplitude rather than
just its absorbtive part (as is the case in the conventional structure
functions).
Because of this it is not possible to give a formal derivation of scaling for
deep inelastic electropion production processes even if one believes that they
are unambiguously sensitive to the light cone behavior of the operator product.
 The origin of this is shown to be related to its behavior near $x\approx 0$.
Investigations, both theoretical and experimental, of these processes is
therefore strongly encouraged.}
\Date{06/94}

\newsec{\underbar{INTRODUCTION}}

\subsec{\underbar{Motivation}}

The  well-known logarithmic scaling violations in  the  structure
functions  of nucleons predicted by asymptotic freedom  played  a
crucial role in establishing QCD as the accepted theory of strong
interactions.  These  predictions,  based as  they  are  on  the
renormalization  group and the operator product expansion of  two
electromagnetic  currents near the light cone,  are on  a  strong
theoretical footing. This is in contrast to the situation in many
other hadronic processes.  For example, a theoretical analysis
 of the Drell-Yan process,  or of
jet production,   requires  some  input  of  unknown  soft   infrared
contributions \ref\rlone{see, for example, J. -P. Cheng and L. -L. Li''Gauge
Theory of Elementary Particle Physics'', (Oxford University Press, Oxford, UK
1984)}.  In spite of this ignorance,  theorists  concerned with
these  processes  have gone ahead with  recipes  for  their calculation,
which often have been very successful. These recipes involve a careful mix
of several ingredients,  some of which such as   asymptotic  freedom
and  perturbation  theory   are   well-understood,  while others, such as
the ingredient describing non-perturbative hadronization effects are
chosen with an eye for the acceptability  of  the  finished product rather
than  for  their theoretical  basis.  Nevertheless this pragmatic
approach is  now taken  to  be sufficiently reliable that it is used  to
estimate backgrounds   in  experiments  searching  for  unusual   or
new phenomena: for example calculations of jet production are used in
searches for potential Higgs candidates.

It is therefore important for theorists to appreciate the lack of
knowledge  which  is  a necessary input  into  such  calculations
involving  perturbative QCD so that experimental results  can  be
used to illuminate the approximations and assumptions used in the recipe.
It  is  in this spirit  that we  consider  here  another hadronic  process
which  is similar in  many  respects  to  the classic deep inelastic
structure function process which underpins perturbative QCD. This is the
process of deep inelastic exclusive pion  production  from nucleons which
each of us worked  on  many years  ago  using  the ideas of current
algebra  and  PCAC.  Our original analyses were clearly not rigorous
but they did lead  to predictions  of scaling which agree with
 experiment albeit at relatively low energies.  We therefore re-analyse this
process from a modern viewpoint in this paper: we derive the scaling laws
and  calculate the expected logarithmic deviations from  scaling.

It  seems to us that it is important to stimulate  more  interest
in  this  and similar processes at this time   because  they  are
amenable  to  experiment in the not-too-distant future  at  HERA,
CEBAF  and  a possible upgrade of SLAC.  Much of what we  say  is
applicable   also   to  other  processes  which   are   described
theoretically  by  an  amplitude which  is  proportional  to  the
product of two currents.  This should therefore include the  deep
inelastic   electroproduction  of  $\rho$'s,   $K$'s   and   most
interestingly real photons: the latter process corresponding to a direct
measurement on the non-forward Compton  amplitude,  albeit with  one
virtual and one real photon.  We will show that if  the predictions   of
scaling  behaviour  are  indeed   verified   by experiment,  important
implications  follow about  the  analytic structure of the amplitudes for
these processes.

In section 1.2 we define the process  we are interested in and the
corresponding kinematic region. In section 2.1 we outline the derivation
of the scaling law as was done originally by two of us over ''twenty'' years
ago \ref\rltwo{N. Dombey
and R. T. Shann, Phys. Lett. \underbar{42B}, 486(1972)}\hskip
1mm\ref\rlthree{G. B. West, ibid \underbar{46B}, 111 (1993)} using Current
Algebra. The resulting
predictions were, in fact, verified by some rather coarse data taken
about that time \ref\rlfour{C.J. Bebek et al., Phys. Rev. Letters
\underbar{30}, 624 (1973)}\hskip 1mm\ref\rlfive{C. J.
Bebek et al. ibid \underbar{34}, 759 (1975)}. In section 2.2 we
present a rederivation of the scaling law using the language of the
parton model as done in the thesis of one of us in the early
1980's \ref\rlsix{A. Calogeracos, D.Phil. Thesis, University of Sussex (1981)}.
  In section 2.3 we calculate logarithmic corrections to the
amplitude using the diagrammatic approach pioneered in the seminal paper
by Altarelli and Parisi \ref\rlseven{G. Altarelli and G. Parisi,
Nucl. Phys.\underbar{B126}, 298 (1977)}.  It should be noted that the end
result of this approach is the integral \future\twentynine \ for the
amplitude {\it M}, which is analogous to the Altarelli-Parisi evolution
equation. This equation can then be used  to predict  the moments of
the amplitude, which in principle can be measured. The validity of the
procedure is subject to some reservations which we will discuss fully.
In section 3 we approach the problem from the point of view of
Operator Product Expansion. We explain why the
prediction for the moment of the amplitude is sensitive to the
analytic properties of the amplitude near x=0. This implies that an
experimental study of deep inelastic pion production from which
these moments can be determined may well yield information on the low
x dependence of the amplitude. In section 4 we discuss the connection
of our results to experimental data and suggest future experiments.

\subsec{Kinematics \& Definitions}

The amplitude that we are going to study is defined as follows:
$$
\eqalignno{M_{\mu}&{}=\langle p^{\prime} \pi | J_{\mu} | p\rangle
&\eqnn\myeqna\myeqna\cr
&{}=(m^{2}_{\pi}-q^{\prime 2})\int d^{4}xe^{iq.x}\langle
p^{\prime}|\theta(x_{0})[J_{\mu} (x), \phi_{\pi}(0)]|p \rangle
&\eqnn\myeqnb\myeqnb\cr
&{}={(m_{\pi}^{2}-q^{\prime 2})\over f_{\pi} m_{\pi}^{2}} \int d^{4}x
e^{iq.x}\langle
p^{\prime}|\theta(x_{0})[J_{\mu}(x), \partial^{\nu}A_{\nu}(0)]|p\rangle
&\eqnn\myeqnc\myeqnc\cr}
$$

Here $J_{\mu}(x)$ is the electromagnetic current, $A_{\mu}(x)$ the axial
current, $m_{\pi}$ the
pion mass, $f_{\pi}$ its decay coupling constant and $\phi_{\pi}(x)$ its field.
 In going from
$\myeqnb$ to $\myeqnc$, the standard PCAC identification has been used:
\eqn\two{\partial_{\mu} A^{\mu}(x)
= f_{\pi} m^{2}_{\pi} \phi_{\pi}(x).} The kinematics are illustrated in fig. 1:
$p$ is the 4-momentum of the
struck target, $p^{\prime}$ its final momentum and $q$ that of the virtual
photon delivered by the scattered
electron; $q^{\prime}$ will be used for the pion 4-momentum.

The relationship to, and generalization from the amplitude probed by measuring
the
conventional structure functions is clear.  In that case one is probing only
the
imaginary part of the forward Compton amplitude whereas in electropion
production one
measures a {\underbar{full}} amplitude which, in general, is
\underbar{non}-forward.
Formally, the difference can be expressed as probing the difference between a
time-ordered, or
retarded product, as in \myeqnc, versus a commutator as in the
structure function case: \eqn\three{W_{\mu\nu} = \int d^{4}x e^{iq.x}\langle
p|[J_{\mu}(x),
J_{\nu}(0)]|p\rangle.} The full forward Compton amplitude is given by
\eqn\four{{\cal J}_{\mu\nu} \equiv \int d^{4}x e^{iq.x}\langle p|T[J_{\mu}(x),
J_{\nu}(0)]|p\rangle} so that $W_{\mu\nu} = {\rm Im}\; \;{\cal J}_{\mu\nu}$.
These are represented by the diagrams in fig. 2.

A further crucial difference between the two cases is, of course, that in
\myeqnc, the
kinematics of real pion production dictates that, even in the deep inelastic
limit
when $q^{2}$ is large, $q^{\prime 2}$ must remain fixed at $m^{2}_{\pi}$; on
the
other hand, in \three, the magnitude of the virtual mass of \underbar{both}
currents is
always large in the deep inelastic limit.  This latter condition ensures that
the
light-cone is unambiguously being probed and so justifies the use of the light
cone
operator product expansion.  In spite of the fact that this is not clearly the
case
in pion production we shall argue below that a short distance operator product
expansion may dominate the process when $q^{2}$ is large.

There is a subtlety in this procedure which is also present in the  standard
forward Compton amplitude case.  The point is that this formalism leads to an
expansion in powers of ${1/x}$, where $x\equiv -q^{2}/2p.q$, and, in the
physical region accessible to real experiments, $|x|<1$.  Such an expansion
therefore
clearly does not converge.  In the structure function case, this potential
problem is
finessed because, there, one is interested in only the imaginary part, as in
\three,
so an analytic continuation from the unphysical large $|x|$ (where the
expansion presumably makes sense) to the physical region can be effected
\ref\rlone\hskip 1mm\ref\rleight{J. Collins, "Renormalization", Cambridge
University Press, Cambridge 1984, Chapter 14}.  Indeed this is why
the results are expressed in the form of moments of the structure functions
rather
than the structure functions themselves.  We shall discuss this in more detail
below.
 However, it is clear from this, that in the pion production case $M_{\mu}$ is
sensitive to a
potentially interesting part of the formalism not readily accessible to the
structure
functions.  Indeed it may well be that because of this ``problem'' pion
electroproduction can cast interesting light on the general small $x$ behavior
as well as
the general assumptions that underly the usual derivation.
Before reviewing the old scaling arguments, however, let us recall the
relationship
between the measured cross-section and the matrix element $M_{\mu}$: this is
best done
in terms of the tensor \eqn\five{T_{\mu\nu} \equiv M_{\mu} M^{\ast}_{\nu}.}
The result is given by: \ref\rlnine{See, e.g., C.W. Akerdof et al., Phys. Rev.
\underbar{163}, 1482
(1967)} \eqn\six{{d^{3}\sigma\over dE^{\prime}d\Omega^{\prime}d\Omega} =
{\alpha \over
2\pi^{2}q^{2}}{E^{\prime}\over E}{(v^{2}-q^{2})^{1/2}\over
1-\epsilon}{d\sigma\over
d\Omega}.}
where $E(E^{\prime})$ is the initial (final) energy of the electron in the
Laboratory
(LAB) system and $\nu$ its energy loss: note that $\nu = E-E^{\prime}=p.q/M$
where
$M$ is the target nucleon mass.  The polarization of the virtual photon is
given by:
\eqn\seven{\epsilon = \lbrack 1 - {2(\nu^{2}-q^{2})\over q^{2}}\tan^{2}{1\over
2}\theta_{e}\rbrack^{-1}}where $\theta_{e}$ is the electron scattering angle in
the
LAB.  The quantity $d\sigma/d\Omega$ represents an equivalent virtual
photoproduction cross-section in the outgoing hadron center-of-mass (CM)
system:
\eqn\eight{\eqalign{{d\sigma\over d\Omega} & = {M^{2}|q^{\prime}|_{{\rm
CM}}\over
16\pi^{2}W^{2}|q|_{\rm CM}}\lbrack {1\over 2}(T_{xx}+T_{yy})+{1\over
2}\epsilon(T_{xx} - T_{yy})\cr
 & - (q^{2}/\nu^{2})\epsilon
T_{zz}+\{-(2q^{2}/\nu^{2})\epsilon(1+\epsilon)\}^{1\over
2}T_{xz}\rbrack\cr}}
Here $W$ is the total $CM$ energy so $W^{2}\equiv
s=(p+q)^{2}=(p^{\prime}+q^{\prime})^{2}$. The $z$-axis is defined to be
coincident with the direction
of $q$ whilst the electrons define the $xy$ plane.  Thus all of the $\phi$
(azimuthal) dependence is
contained in $(T_{xx}-T_{yy})\sim\cos 2 \phi$ and $T_{xz}\sim\cos \phi$.  In
what follows we shall
limit ourselves to the case where the particle spins are unobserved.

Finally, it is worth noting that in the deep inelastic limit $(q^{2}\rightarrow
- \infty)$, the square of
the momentum transfer
\eqn\nine{t\equiv \Delta^{2}\equiv (q^{\prime}-q)^{2} = (p^{\prime}-p)^{2}} is
constrained, in the
physical region, to lie between $-2\nu$ and
\eqn\ten{t_{{\rm min}} \approx - {x^{2}M^{2}\over (1-x)}.}
Typically the limit we will be considering keeps $t$ and $x$ fixed (and finite)
with $q^{2} \rightarrow -\infty$.
Thus $x$ must not be too close to unity.  Furthermore, the region of interest
is predominantly forward
scattering in the $\pi N$ CM system.  In what follows it is convenient to write
\eqn\ninea{ q_{\mu} = En_{\mu} +\Delta_{\mu}} where $n^{2}$ is a null-vector,
i.e. $n^{2}=0$.
Thus \eqn\nineb{q^{2} = 2E(n.\Delta)+t} and
\eqn\ninec{\eqalign{x = {-\lbrack
2E(n\cdot\Delta)+t\rbrack \over(2\lbrack E(n\cdot p)+\Delta \cdot p\rbrack)}}.}
 The scaling limit can
then be realized by taking $E\rightarrow \infty$ with both $x \approx -
(n.p)/(n.\Delta )$ and $t$ fixed.

\newsec{\underbar{SCALING LAW}}
\subsec{\underbar{Current Algebra}}

Scaling laws for $T_{\mu \nu}$ can be derived using a current algebra approach
augmented by some heuristic
assumption about the light cone behavior of the commutator.  This can be
checked in perturbation theory and
justified by the operator product expansion as sketched below.  We begin by
setting $q^{\prime 2}=0$ in
which case
\eqn\elevena{\eqalign{& f_{\pi} M_{\mu} = C_{\mu \nu} q^{\prime \nu} +
E_{\mu}\cr}}
where
\eqn\elevenb{\eqalign{& C_{\mu\nu} \equiv i \int d^{4}x\; {\rm exp}\; (iq\cdot
x)\langle
p^{\prime}|\theta(x_{0})\lbrack J_{\mu}(x), A_{\nu}(0)\rbrack |p\rangle\cr}}
and
\eqn\elevenc{\eqalign{& E_{\mu}\equiv \int d^{4}x {\rm exp}(iq.x)\langle
p^{\prime}|\delta(x_{0})\lbrack
J_{\mu}(x), A_{0}(0)\rbrack |p\rangle \cr}.}

Using the usual SU(2) x SU(2) current algebra
\eqn\twelve{\delta (x_{0})\lbrack A^{i}_{0}(x), J_{\mu}(0)\rbrack
=i\epsilon^{i3k}A^{k}_{\mu}(0)\delta^{4}(x)} we immediately get that $E_{\mu}$
is independent of $E$; (it
depends only on $\Delta$ and, from its usual parametrization, we get the
well-known axial vector and induced
pseudoscalar form factors of the nucleon).

The scaling result we want to show is that $C_{\mu\nu} q^{\prime \nu}$ is also
independent of $E$.  We
shall first sketch the derivation of this result based on the spacetime
behavior of the current
commutators.
In what follows it is convenient to introduce standard light-cone coordinates
for a four-vector
$a^{\mu}$ as follows:
\eqn\thirteen{a_{\pm} = {\sqrt{2}\over 2}(a_{0} \pm a_{z});\;\; {\bf a}_{\perp}
= (a_{x} , a_{y}).}
Then the scaling limit is equivalent to $q_{-}\approx \sqrt{2}\nu \rightarrow
\infty$ with $q_{+}\approx
\sqrt{2}x$ fixed.

Causality allows us, at least naively, to replace $\theta (x_{0})$ in (2.2) by
$\theta(x_{+})$ in which case
the asymptotic behavior of $C_{\mu\nu}$ is given by
\eqn\fourteen{C_{\mu\nu}\approx -{1\over q_{-}} \int d^{4}x \exp (iq\cdot x)
\langle p^{\prime}|\delta
(x_{+})\lbrack J_{\mu} (x), A^{i}_{\nu}(0)\rbrack |\rangle}
Consequently,
\eqn\fourteen{q^{\prime \nu}C_{\mu\nu} \sim q_{-}C_{\mu+}\sim - \int d^{4}x
\exp(iq\cdot x)\langle
p^{\prime}|\delta(x_{+})\lbrack J_{\mu} (x), A^{i}_{+}(0)\rbrack | p \rangle.}
The commutator in (2.7) can be expressed in the form
\eqn\fifteen{\lbrack J_{\mu}(x), A^{i}_{+}(0)\rbrack \delta(x_{+}) =
\tilde{A}_{\mu}^{i}(x_{-})\delta^{2}(x_{\perp})} where $\tilde{A}_{\mu}(x_{-})$
is, in general,
model-dependent.

In QCD, canonical commutation relations lead to
\eqn\sixteen{\tilde{A}^{i}_{\mu}(x_{-}) = A^{i}_{\mu}(0)  \delta(x_{-}) +
B^{i}_{\mu}(x_{-})}
where $B^{i}_{\mu}(x_{-})$ is an unknown (typically bilinear non-local)
operator which is non-singular at
$x_{-}\approx 0.$
We conclude then that, in the scaling limit,
\eqn\seventeen{q^{\prime \nu}C_{\mu\nu}\sim\int dx_{-}\exp(iq_{+}x_{-})\langle
p^{\prime}|B_{\mu}(x_{-})|p\rangle.}
This is the desired result since it shows that $M_{\mu}$ is independent of $E$
and is only a function of
$x$ (through $q_{+}$) and $\Delta$.

This result straightforwardly translates into the following scaling constraints
on the components of
$T_{\mu\nu}$ occurring in the measured cross-section, $\six$:
\eqn\eighteena{\eqalign{
& {1\over 2}(T_{xx}+T_{yy})\rightarrow F_{1}(x,t)(k^{2}_{x}+k^{2}_{y})\cr
& {1\over 2}(T_{xx}-T_{yy})\rightarrow {1\over
2}F_{2}(x,t)(k^{2}_{x}-k^{2}_{y})\cr
& T_{zz}\rightarrow F_{1}(x,t)-F_{2}(x,t)\Delta^{2}_{z}\cr
{\rm and}\;
& T_{xz}\rightarrow -F_{2}(x,t)k_{x}\Delta_{z}\cr}} where $\Delta_{z}$ (as
expressed
in the LAB)$\approx -({1\over 2}t+M\omega)$ and the $F_{i}$ are Lorentz
scalars.
\subsec{\underbar{Parton Model}}

We now turn to the treatment of this problem in terms of the quark-parton
model.  We assume that the two
currents $J_{\mu}$ and $A_{\nu}$ interact successively with the same quark
while the rest act as
spectators.  Without QCD corrections the amplitude $C_{\mu\nu}$ is given by the
sum of the two diagrams in
fig. 3.
This is analogous to the usual parton model treatment of the forward Compton
scattering.

To calculate these diagrams we assume that the struck quark carries a fraction
$\eta$ of the momentum $p$ of
the hadron, in the infinite momentum frame. Immediately after absorbing the
photon the virtual quark has momentum $(\eta p +
q)$, and hence it is highly off-shell. This is the basic reason for considering
that the parton model is applicable here.   The final quark has momentum $(\eta
p +
\Delta)$ .  Consider now the diagram in fig. 3(a).  Its contribution is
$(\tau_{i}$ are isospin
matrices and $\tilde{Q}$ is the generator corresponding to the electric
charge):
\eqn\nineteen{\eqalign{M_{(a)}& = - {\tau_{i}\over
2}\tilde{Q}{\tilde{\psi}_{p{\prime}}(\eta p +
\Delta )\gamma_{5}\gamma_{\nu}(\eta {\not{p}} + E {\not{n}} +
{\not{\Delta}})\gamma_{\nu}\psi_{p}(\eta
p) \over 2E\lbrack \eta (n.p) + (n.\Delta )\rbrack} q^{\prime}_{\nu}\cr
& = - {\tau_{i}\over 2}\tilde{Q}{\tilde{\psi}_{p^{\prime}}(\eta p + \Delta
)\gamma_{5} {\not{n}} (\eta {\not{p}}
+ {\not{\Delta}})\gamma_{\mu}\psi_{p} (\eta p)\over 2\lbrack \eta (n.p) +
(n.\Delta )\rbrack}\cr}}
To this has to be added the contribution from the crossed graph shown in fig.
3(b).
A sum over the various types of quarks has been suppressed.  $\psi_{p}(k)$
represents the amplitude, or
wave function, for finding a quark of a particular type carrying momentum $k$
inside a nucleon moving with
momentum $p$.  The complete matrix element requires an integration over $\eta$,
consistent with the
requirement that $(\eta p + q)^{2} > 0$ i.e. $\eta > x$.  Schematically, the
parton contribution is thus
given by
\eqn\twenty{{\cal M}^{0}_{\mu} = \int^{1}_{x}d\eta
\overline{\psi}_{p\prime}(\eta p + \Delta)M_{\mu}(\eta ,
p, \Delta) \psi_{p}(\eta p).}
  The matrix ${\cal M}_{\mu}$ can be read off from (2.13) with an additional
contribution coming from the
crossed graph.  This shows explicitly that, when $E$ is large, $M$ depends only
on $t$ and $x$.

Note, incidentally, that for the parton model description of the conventional
structure function, $\Delta =
0$ and only the imaginary part of ${\cal M}$, i.e. its delta-function
contribution, is required. In
that case the integral reduces to $x f(x)$, where $f(\eta)\equiv
\overline{\psi}_{p}(\eta p)
\psi_{p}(\eta p)$, the probability for finding a quark with fraction $\eta$ of
the total momentum.
It is worth remarking that in the analogous kinematic configuration here, where
$\Delta \approx 0$,
the full Born amplitude reduces to \eqn\twentyone{M_{\mu}\approx \lbrack Q,
{\tau^{i}\over 2}\rbrack
(n.p) p_{\mu} .} \subsec{\underbar{Leading Logarithmic Corrections}}

In this section we sketch a computation of the leading logarithmic corrections
to the parton model result.
Rather than give a complete detailed description we present here only the
salient features for the simpler
and more limited kinematic situation where $\Delta \approx 0$.  The point is
that a subgraph which gives a
logarithmic contribution $\sim\ln \Delta^{2}$ for $\Delta \neq 0$ obviously is
singular in the $\Delta
\rightarrow 0$ limit: hence the same diagrams (i.e. ladders and self-energy
insertions to external legs)
that give the leading logarithmic corrections in the usual deep inelastic
scattering case ($\Delta = 0$)
also give the leading logarithmic corrections in the $\Delta\neq 0$ case.

For ease of presentation, we shall from here on use $p$ to denote the momentum
of the initial quark rather than of the initial hadron.  The initial quark is
taken to be off-shell by an amount comparable to the inverse of the
confinement radius of the nucleon.  Since the momentum transfer $\Delta$ is
typically of the same order of
magnitude, we make no distinction between $\ln(-q^{2}/p^{2})$ and $\ln
(-q^{2}/\Delta^{2})$.  When,
for example, we end up with a logarithmic integration of the form $\int
dk^{2}/(k^{2} + \Delta^{2})$
we shall be free to take the $\Delta \rightarrow 0$ limit and write it in the
form $\int_{p^{2}}
dk^{2}/k^{2}$.  Although the final scattered quark is slightly off-shell, the
$\Delta\rightarrow 0$
limit enables us to take the $\gamma$
 matrices between on-shell spinors.  It is important, of course, to take the
$\Delta \rightarrow 0$
limit \underbar{after} having secured that the final integration is
logarithmic.

It is convenient to work in the light cone gauge where the propagator for a
gluon of momentum $k$ is
\eqn\twentytwoa{\eqalign{G_{\mu \nu}(k) & = {D_{\mu \nu}(k)\over k^{2} +
i\epsilon}\cr}}
with
\eqn\twentytwob{\eqalign{D_{\mu \nu}(k) & = g_{\mu \nu} -
{k_{\mu}c_{\nu}+k_{\nu}c_{\mu}\over
k.c}\cr}.} In such a gauge the only diagrams (apart from self energy parts)
giving leading logarithmic
corrections are the ladder diagrams of fig. 4(a).  It may be mentioned, in
particular, that diagrams of
the type of fig. 4(b) do not give leading logarithmic contributions.
Incidentally, had we followed the
operator product expansion approach these diagrams would correspond to
contributions from gauge
non-invariant operators.

Turning now to the calculation of the ladder diagrams we show that a familiar
picture emerges: there is
strong ordering in the momentum flowing through the ladder and an evolution
equation can be derived.  We
first examine the contribution from the lowest order diagram (fig. 5).  Using a
Sudakov parametrization
for the quark momentum $k$
\eqn\twentythree{k =\alpha c + \beta p + k_{\perp}}
\eqn\twentythreea{d^{4}k = {s\over 2} d\alpha d \beta d^{2}k^{2}_{\perp}}
(where $s=-{q^{2}/x}$).  We obtain:

\eqn\twentyfour{M_{1}=-{\tau i \over 2}\tilde{Q} {\alpha_{s}C_{F}\over
4\pi^{3}}{s\over 4}\int
dk_{\perp}^{2}d\alpha d\beta}
$${\gamma_{\sigma}(\not{k}+\not{\Delta})\not{n}(\not{k}+\not{\Delta})
\gamma_{\mu}\not{k}\gamma_{\rho}\gamma_{5}\over
\lbrack (k+E_{n}+\Delta)^{2}+i\epsilon\rbrack (k^{2}+i\epsilon)\lbrack
(k+\Delta)^{2}+i\epsilon\rbrack\lbrack
(p-k)^{2}+i\epsilon\rbrack}D_{\rho\sigma}
$$

Initial and final spinors have been suppressed.  Note that the
amplitude is color singlet in the t-channel and that, in terms of Sudakov
variables,
\eqn\twentyfive{k^{2}=\alpha\beta s - k^{2}_{\perp},\;\; (p-k)^{2} = - \alpha
(1-\beta )s-k^{2}_{\perp},\;\;
(k+\Delta)^{2}=\alpha \beta s - k^{2}_{\perp} -\alpha s x.} We first perform
the $\alpha$ integration.  In
the region $x < \beta < 1$ there is one pole at $\alpha =
-k^{2}_{\perp}/(1-\beta)s$ due to $(p-k)^{2}$
lying below the real axis whereas in the region $0<\beta<x$ there is one pole
at $\alpha =
k_{\perp}^{2}/\beta s$ due to $k^{2}$ lying above it.  In both cases we close
the $\alpha$ contour so as to
pick the contributions from those poles.  In the regions $\beta < 0$ and $\beta
> 1$ all the poles with
respect to $\alpha$ lie on one side of the real axis and can be avoided (we do
not take $(k + En + \Delta
)^{2}$ into account since it will be combined with the next element of the
ladder).
Observe that the leading logarithms come from the wide range of integration
$\mu^{2}\leq k_{\perp}^{2}\ll s$ where $\mu$ is an arbitrary renormalization
scale.  The
parameter $\beta$ is finite (typically of order $x$) whereas $\alpha \sim -
k_{\perp}^{2}/s$ is
small.  Hence the logarithms come, as expected, from the colinear
configuration.

After the $\alpha$ integration the denominator behaves like
$(k_{\perp}^{2})^{2}$, so we have to extract
one $k_{\perp}^{2}$ from the numerator if the final integration is to be
logarithmic.  Having done this we
can pass to the collinear configuration $k=\beta p,\; \alpha \approx 0$.  As
already remarked we
shall also set $\Delta \approx 0$ and take the $\gamma$ matrices between
on-shell spinors $u(p)$.  One
might worry whether in the limit $\Delta \rightarrow 0$ we lose logarithms
multiplied by
$q.\Delta/q^{2}$.  There is no such danger since we parametrize everything from
the start in terms of
vectors $n, p$ and $\Delta$.  Then $q.\Delta/q^{2}\sim
E(n.\Delta)/2E(n.\Delta)\sim {1\over 2}$.  There
will remain a factor $\gamma_{5}{\not{n}}{\not{k}}\gamma_{\mu}$ from the
numerator which will combine with
$(k+En+\Delta)^{2}$ from the denominator to form the Born term $M_{B}(\beta p,
k_{\perp}^{2})$.  Note
finally that the logarithmic contribution comes from the region $\beta > x$, so
that the propagating
quark line remains highly off-shell.

The numerator in the integrand has the form (apart from the $\gamma_{5}$):
\eqn\twentysix{\gamma_{\sigma}(\not{k} +
\not{\Delta})\not{n}(\not{k}+\not{\Delta})\gamma_{\mu}\not{k}
\gamma_{\rho}\big\lbrace
g_{\rho\gamma}-{c_{\rho}(p-k)_{\sigma}+c_{\sigma}(p-k)_{\rho}\over
c.(p-k)}\big\rbrace.}
After some algebra this can be reduced (in the $\Delta \approx 0$ limit) to a
familiar form for the
leading lowest order correction:
\eqn\twentyseven
{M_{1}(x,q^{2})=2C_{F}\int^{-q^{2}}_{\mu^{2}}{dk^{2}_{\perp}\over
k^{2}_{\perp}}{\alpha s\over
4\pi}\int^{1}_{x}d\beta \lbrace 1-\beta + {2\beta\over 1-\beta}\rbrace
M_{B}(x/\beta , \mu^{2})}
Iterating this an arbitrary number of times leads to an evolution equation:
\eqn\twentyeight{M(x,q^{2})=M(x,\mu^{2})+2C_{F}\int^{-q^{2}}_{\mu^{2}}
{dk^{2}_{\perp}\over
k^{2}_{\perp}}{\alpha_{s}\over 4\pi}\int^{1}_{x}d\beta {1+\beta^{2}\over
1-\beta} M(x/\beta ,
k^{2}_{\perp}).}

Up to now we have considered skeleton graphs only.  When we dress the ladder
with vertex and self-energy
corrections further leading logarithmic contributions coming from the
ultraviolet region are induced.
These can be taken into account simply by replacing the ``bare'' coupling
constant $\alpha_{s}$ in
\twentyeight \hskip 1mm by the running coupling constant
$\alpha_{s}(k^{2}_{\perp})\approx 4\pi/(\beta \log k^{2}_{\perp})$ where $\beta
= (11 -2n_{f}/3)$ ($n_{f}$
being the number of flavors).  In addition, the second term in \twentyeight
\hskip 1mm
must be multiplied by the quark wavefunction renormalization constant $Z_{F}$
in order to cancel the
infrared divergences from the soft gluon region and get a gauge invariant
result.  $Z_{F}$ in the
light-cone gauge has been calculated in a number of places.  The final result
is
\eqn\twentynine{M(x,q^{2})=M(x,\mu^{2})+2C_{F}\int^{-q^{2}}_{\mu^{2}}
{dk^{2}_{\perp}\over
k^{2}_{\perp}}{\alpha_{s}(k^{2}_{\perp})\over 4\pi}\int^{1}_{x}d\beta P(\beta)
M(x/\beta ,
k^{2}_{\perp})} where,  \eqn\thirty{P(\beta)={1+\beta^{2}\over
1-\beta}-\delta(1-\beta)\int^{1}_{0}dx{1+x^{2}\over 1-x}.}

Equation \twentynine \ is somewhat difficult to handle from the
phenomenological
point of view.  If for the moment we disregard the subtleties regarding the low
$x$ dependence of $M$ which will be discussed in the following section, then
we can disentangle $M$ in \twentynine \ by taking moments.  Defining
\eqn\thirtyone{M_{n}(q^{2}) = \int^{1}_{0} dx x^{n}M(x,q^{2})}
we get \eqn\thirtytwo{M_{n}(q^{2})\sim
 \Bigl( \ln {q^{2}\over \mu^{2}}\Bigr)^{dn}}
where \eqn\thirtythree{d_{n}={C_{F}\over \beta}\Bigl(1 + 4\sum^{n+2}_{j=2}
{1\over j}{2\over (n+2) (n+3)}\Bigr)}

\newsec{\underbar{OPERATOR PRODUCT EXPANSION}}

In this section we discuss an operator product expansion (OPE) analysis of this
amplitude.  As
already intimated, there are some subtleties that prohibit a straightforward
prediction for its asymptotic behavior.
Before discussing this, however, it is   worth reviewing briefly the standard
treatment of the
conventional structure functions from this viewpoint.  In that case the use of
perturbation theory
to determine the large $q^{2}$ behavior can be justified from the application
of the OPE to the
light cone expansion of ${\cal J}_{\mu \nu}$, \four .

Explicitly, the time-ordered product of the two currents can be Taylor expanded
around
$x^{2}\approx 0$ in terms of a complete set of operators
$O^{\mu_{1}....\mu_{n}}_{mn}$:
\eqn\thirtyfour{T\lbrack J(x) J(0)\rbrack \approx \sum_{m , n}
c_{m}(x^{2})x_{\mu_{1}}....x_{\mu_{n}}O^{\mu_{1}....\mu_{n}}_{mn} (0).}
[For ease of presentation, the currents are here taken to be scalar].
Equivalently, its
Fourier transform is given by \eqn\thirtyfive{\int d^{4}x e^{iq.x}T\lbrack J(x)
J(0)]\approx
\sum_{m, n} C_{mn}(q^{2}){q_{\mu_{1}}...q_{\mu_{n}}\over ({1\over 2}
q^{2})^{n}}O^{\mu_{1}....\mu_{n}}_{mn}} where
\eqn\thirtysix{C_{mn}(q^{2})\equiv \big(
-{i\partial\over \partial \ln q^{2}}\big)^{n} \int d^{4}xe^{iq.x}
c_{m}(x^{2}).}

On dimensional grounds the $C_{mn}(q^{2})$ behave, up to logarithms, like
$(q^{2})^{-d_{c}}$ for
large $q^{2}$ where $d_{c} = 2d_{J}-4-(d_{0}-n)$ is the dimension of $C$,
$d_{J}$ that of $J(x)$
and $d_{0}$ that of the operator $O^{\mu_{1}....\mu_{n}}_{mn}$.  This is, of
course, the origin
of the observation that the asymptotic behavior is controlled by the operator
having the lowest
twist $\tau_{0}\equiv d_{0}-n$.  Notice that these equations are all properties
of the
current operators (i.e., in the case of interest here, the pion and the virtual
photon) and do not
depend on the target state.  For forward scattering we require the ground state
target matrix
elements \eqn\thirtyseven{\langle p|O^{\mu_{1}....\mu_{n}}_{mn}|p\rangle =
A_{mn}p_{\mu_{1}}....p_{\mu_{n}} +
B_{mn}g_{\mu_{1}\mu_{2}}p_{\mu_{3}}....p_{\mu_{n}}+\cdot}

The $A_{mn}, \; B_{mn}$ etc. are simply numbers characterizing the target.  In
the contraction
of this with \thirtyfive \ it is clear that, in the Bjorken limit, terms
involving the $A_{mn}$
dominate: one thereby obtains
 \eqn\thirtyeighta{\eqalign{{\cal J}(x,q^{2})& \equiv \int d^{4} x
e^{iq.x}\langle p|T\lbrack J(x)J(0)\rbrack|p\rangle\cr&{}\approx \sum_{m , n}
{A_{mn}C_{mn}(q^{2})\over x^{n}}\cr}}
The large $q^{2}$ behavior of the $C_{mn}(q^{2})$ can be determined from the
renormalization group
using the asymptotic freedom property of QCD.  Typically, for the leading twist
operator, the
$C_{mn}$ are dimensionless and behave like $(\ln\; q^{2})^{-a_{mn}}$ where
$a_{mn}$ is determined
by the anomalous dimensions of the $O_{mn}$.  Implications for the structure
functions, which are
the absorptive part of ${\cal J}$, can be obtained using the standard analytic
properties of
${\cal J}$.  This leads to the well-known result relating the moments of $W$ to
$C_{mn}(q^{2})$:

\eqn\thirtyninea{\eqalign{M_{n}(q^{2}) & \equiv \int^{1}_{0}dx x^{n-2}[\nu W (x
, q^{2})]\cr
&{}\approx \sum_{m} A_{mn}C_{mn}(q^{2})\cr}}

The sum over $m$ is, of course, finite and typically contains only a rather
small number of
terms.  QCD therefore gives a specific prediction for the $q^{2}$-dependence of
each moment and it
is this that has been successfully checked against experiment \rlone.

Now, suppose that experiments could be performed that directly measure the
large-$q^{2}$ behavior
of the full amplitude ${\cal J}(x , q^{2})$.  What is the QCD prediction for
this?  One immediately
sees the difficulty: the expansion, \thirtyeighta \ , presumably only makes
sense for $|x|>1$ and
this is outside of the physical region.  Indeed the analytic continuation to
$|x|<1$ ultimately leads
to the moment equations, \thirtyninea \ . Ideally, one would like to have a
complementary expansion
valid for $|x|<1$; this would require knowledge of the analytic structure near
$x \simeq 0$ which,
unfortunately is not reliably determined by the RG.  Naively, one could proceed
with ${\cal J}$ just
as one proceeded with the $M_{n}$; i.e., simply take $q^{2}\rightarrow \infty$
in  \thirtyninea \
and pick out the dominant $C_{mn}(q^{2})$ as determined by the smallest
anomalous dimension.  In the singlet case, for example, the conservation of the
stress-energy tensor means that it has
no anomalous dimension and so $M_{2}(q^{2})$ asymptotically approaches a
constant.  This, in turn,
means that the leading behavior of the $T_{1 , 2}(q^{2} , x)$, the two
conventional scalar
amplitudes occurring in the decomposition of ${\cal J}_{\mu \nu}$, is given by
\eqn\thirtynine{T_{1}(q^{2} , x) \approx {T_{2}(q^{2} , x)\over 2x}\approx
{<Q^{2}>\over \pi q^{2}
x}\Big( {3n_{f}\over 16 + 3n_{f}}\Big)}

Now let us examine the extension of this to the non-forward case. Eqns.
\thirtyfour \ - \thirtysix \
remain valid since they are properties of the currents and the expansion
\thirtyfour \ is supposed to be in terms of a complete set of operators;
\thirtyseven \ however, clearly needs to be
generalized. This can be straightforwardly accomplished by writing:
\eqn\fortya{\eqalign{\langle p^{\prime}|O^{\mu_{1}\cdots \mu_{n}}_{mn}|p\rangle
& =
\sum^{n}_{k=0}[A_{mnk}(t) p_{\mu_{1}} \cdots
p_{\mu_{k}}\Delta_{\mu_{k+1}}\cdots \Delta_{\mu_{n}}\cr
 & + B_{mnk}(t) g_{\mu_{1}\mu_{2}}p_{\mu_{3}}\cdots
p_{\mu_{k}}\Delta_{\mu_{k+1}}\cdots
\Delta_{\mu_{n}} + \cdots]\cr}}

Clearly $A_{mnn}(0) = A_{mn}$ and $B_{mnn}(0) = B_{mn}$.  When contracting this
with \thirtyfive \
we shall need the quantity:
\eqn\fortyone{{2\Delta . q\over -q^{2}} = 1 + {t \over q^{2}}.}
It is, therefore, the wider set of coefficients $A_{mnk}(t)$ that dominate the
asymptotic
behavior:  \thirtyeighta\ is thereby generalized to
\eqn\fortytwo{{\cal J}(x, q^{2}, t) \approx \sum^{\infty}_{n=0}
\sum^{}_{m}\sum^{n}_{k=0}
{A_{mnk}(t)\tilde{C}_{mn}(q^{2})\over x^{k}}.}

The $\tilde{C}_{mn}(q^{2})$ are the coefficients appropriate to the axial
current case of interest
here, as expressed in \myeqnc \ and \elevenb \ , and are the analogs of the
$C_{mn}(q^{2})$ of
\thirtyeighta \ . They, too, generally fall with $q^{2}$ like powers of
$\ln(q^{2})$ determined by the
appropriate anomalous dimension.  PCAC ensures that, in $M_{\mu}$, there is an
operator with
vanishing anomalous dimension so that it becomes a function of $q^{2}$ and $t$
only.  In any case,
the corrections to this will, as usual, be powers of $\ln (-q^{2})$.  As
already explained it is not
possible, beyond this, to give the precise prediction for the large $q^{2}$
behavior without
summing the series.

 Finally, it should be noted that the result expressed in  \fortytwo \ is
clearly not valid
unless $t\ll q^{2}$, which means that $x$ must not be too close to $1$.  On the
other hand, probing
scaling and its violation should shed some light on the $x\approx 0$ region: if
the predictions of this paper are experimentally verified it means that the
relevant amplitudes are smooth in the region of small $x$ where  it would
appear that the operator product expansion  breaks down.

\newsec{\underbar{COMPARISON WITH EXPERIMENT}}

In this Section we begin by  reviewing the connection of our results to the
existing experimental
data. At present, the only such data is for the inclusive reaction (and this
was taken over 20
years ago at Cornell). It is possible, however, to extend the above arguments
to this case provided the
mass of the final hadronic ``target'' state $(W^{\prime})$ remains relatively
small. The main
difference is that the scaling function will now depend on $W^{\prime}$ in
addition to $x$ and $t$
so that, instead of  \eqn\fortythree{s^{2}{d \sigma\over dt} \approx F(x, t)}
which is the result implied by  \eighteena \ for the purely
exclusive case, one now expects for the inclusive
\eqn\fortyfour{s^{2}{d^{2}\sigma \over dt
d{W^{\prime}}^{2}} \approx F(x, t, {W^{\prime}}^{2}).}

In the Cornell experiment, $d^{2}\sigma / dt d{W^{\prime}}^{2}$ was measured at
two different values of
$\sqrt{s} (2.66$ and $3.14 GeV)$ but at the {\it same} value of $x$.  The data
was
averaged over $\theta$ and $\phi$.  The scaling result,  \fortyfour \ , implies
that the
spectra, when plotted as a function of $W^{\prime}$, should be identical apart
from a normalization
factor  $(3.14/2.66)^{4} \approx 2.41.$  The data, as can be readily seen in
fig. 6 are in
remarkably good agreement with this prediction.

These data are also presented in terms of the transverse momentum $P_{\perp},$
the transverse momentum
of the pion relative to the direction of the incoming virtual photon and of a
variable $x^{\prime}$
which depends on the longitudinal momentum of the pion:
\eqn\fortyfive{x^{\prime} = {P_{11}\over {(P^{2}_{\rm
max}-P^{2}_{\perp})}^{1\over 2}}}  Here
$P_{11}$ is the pion momentum along the direction of the virtual photon and
$P_{\rm max}$ is
the maximum pion momentum.  In fig. 7
\eqn\fortysix{{E\over \sigma_{\rm tot}} {d^{3}\sigma \over dP^{3}}} is plotted
as a function
of $P_{\perp}^{2}$ at the two values of $W$ mentioned previously and at two
different values of
$x^{\prime}$.  The straight lines are fits of the form $A{\rm
exp}(-BP^{2}_{\perp})$.  The
similarity of the spectra suggests that the $P^{2}_{\perp}$ distribution (for
fixed $x$) does
not depend on $q^{2}$, again in striking agreement with the scaling argument.

It is clearly important for new experiments to be carried out at HERA on deep
inelastic pion electroproduction at high energies check whether scaling
continues to hold subject to the logarithmic violations which follow from
\twentynine \. Furthermore the arguments of this paper do not only apply to
deep
inelastic pion electroproduction. A similar argument could be made for deep
inelastic electroproduction of any particle which couples to a nucleon
by means of a local current operator. This would include
deep inelastic electroproduction of real photons, or of lepton pairs, or of
$\rho$'s, or $K$'s, or $\psi$'s, or $\Upsilon$'s, for example. It would be
especially interesting to check whether the amplitude for the  production of
each of these particles scales in the same way, or whether processes which
involve heavy quarks are different.

\newsec{\underbar{CONCLUSION}}

In this paper we have shown how scaling laws for deep inelastic electropion
production  derived on rather general grounds from QCD-inspired current
algebra,  are manifested in QCD perturbation theory.  Leading logarithmic
corrections are calculated and an evolution equation for the amplitude
derived.  These are quite similar in character to the well-known ones occurring
in the conventional Compton amplitude but have the advantage that the
predictions for the full  amplitude are, in this case, amenable to experiment.
In the Compton case,  only the imaginary parts (the conventional deep inelastic
structure functions) are, in practice, measurable.   However, for the
full amplitude we show that, contrary to one's naive expectation, the usual
deviations from scaling derived from an operator product  expansion analysis
do not lead to a well-defined prediction  in
 the physical region.  Thus, unlike the structure function case, the QCD
perturbation theory result cannot be ``rigorously'' justified from asymptotic
freedom.  The reason for this can be traced back to the behavior of the
amplitude near $x\approx 0$; the OPE leads to an expansion in $1/x$ which
cannot converge for a physical process.  The conventional moment equations
for the structure functions which exploit the known analytic properties of
the amplitude are precisely designed to circumvent this difficulty.  Thus,
observation of the scaling laws and their violation for the {\underbar {full}}
amplitude can potentially shed light on the small $x$ behavior and help clarify
just how far one can push results based on QCD perturbation theory.

With renewed interest in such problems stimulated by recent HERA results and
 the potential of detailed data from CEBAF (albeit at relatively low energies)
we feel that it is important to examine processes such as these that are
natural extensions of the canonical structure functions.  It should be
stressed that these processes should also be viewed as yielding complementary
data on the quark-gluon structure of the nucleon. In future work we intend to
explore this
aspect of the problem in more detail; meanwhile, the main thrust of this paper
is
motivated by the desire  to rekindle interest in such problems. In fact  it was
partly stimulated by a query  along these lines from the experimentalist Bogdan
Povh of
the University of Heidelberg; one of us (GBW) would like to thank him for
his original inquiry.
\eject
\newsec{\underbar{FIGURES}}
\input psfig
\centerline{\psfig{file=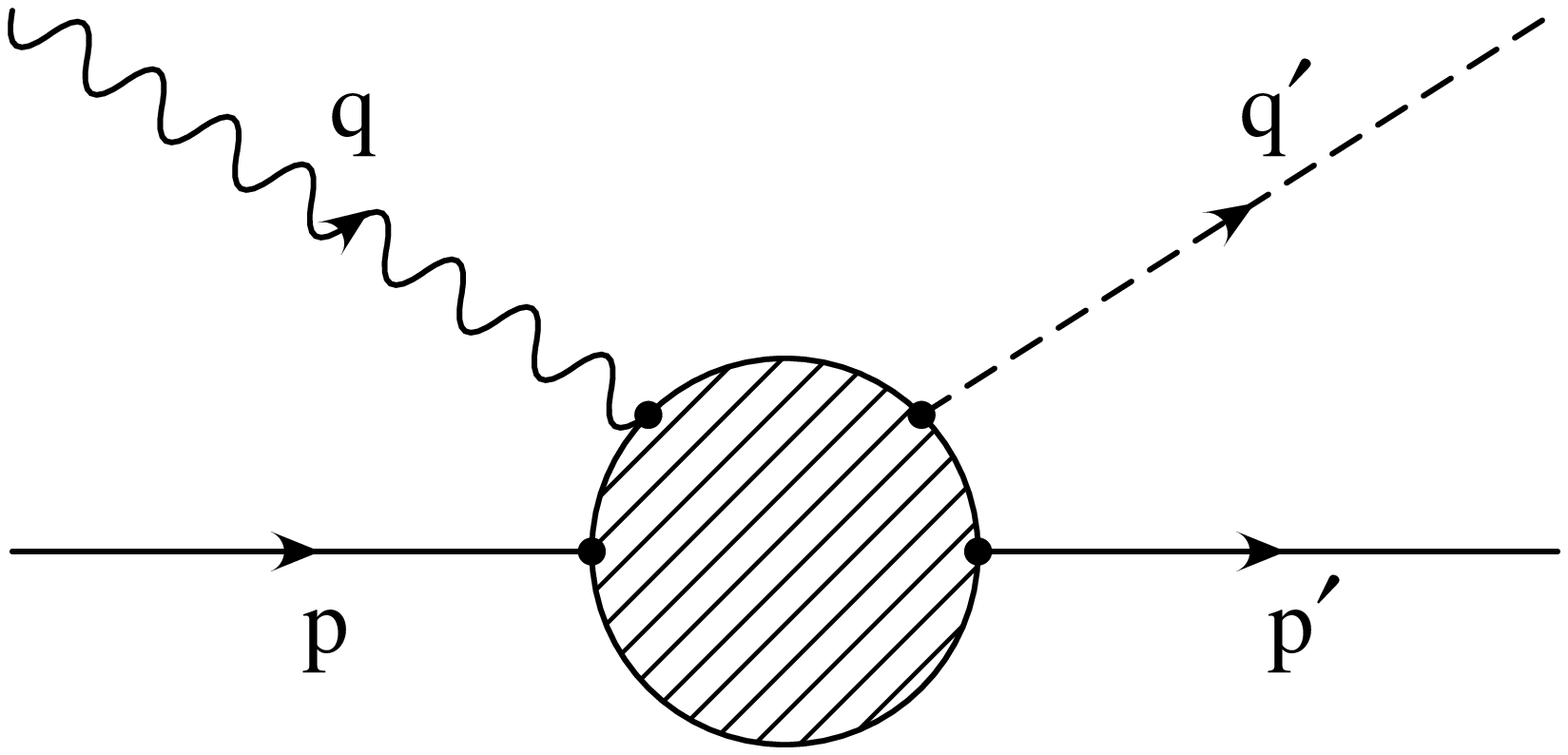,height=3.4in}} \vskip4pt\centerline{Fig. 1}
\hangindent\parindent{General electropion production amplitude defined in
\myeqna \
showing the kinematics of the external particles.}
\vskip3ptplus1fil\vbox{\centerline{\psfig{file=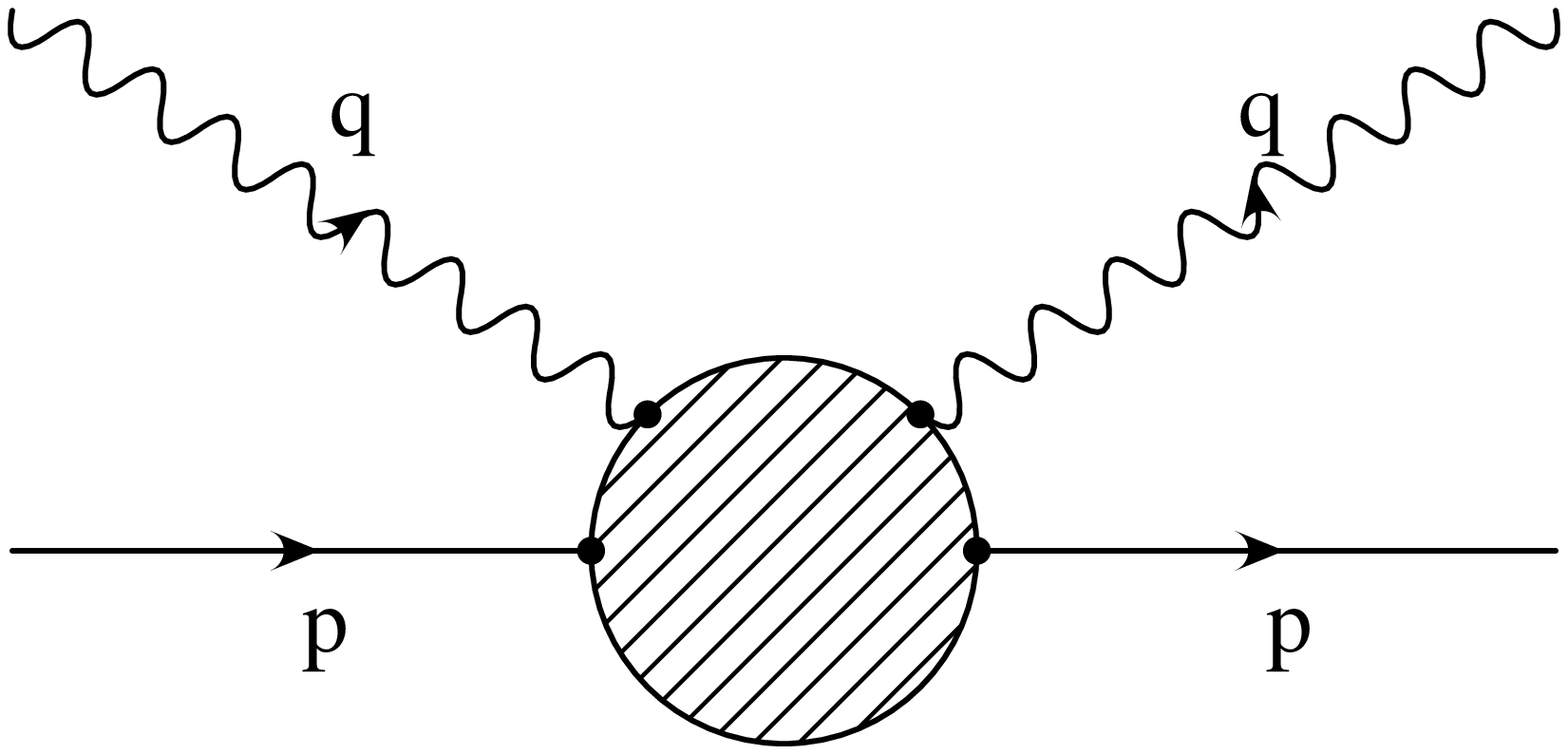,height=3.4in}}
\vskip4pt\centerline{Fig. 2}
\hangindent\parindent{The forward Compton amplitude defined in \four \ ; its
imaginary part defines the conventional structure functions, \three \ .}}
\eject
\vbox{\centerline{\psfig{file=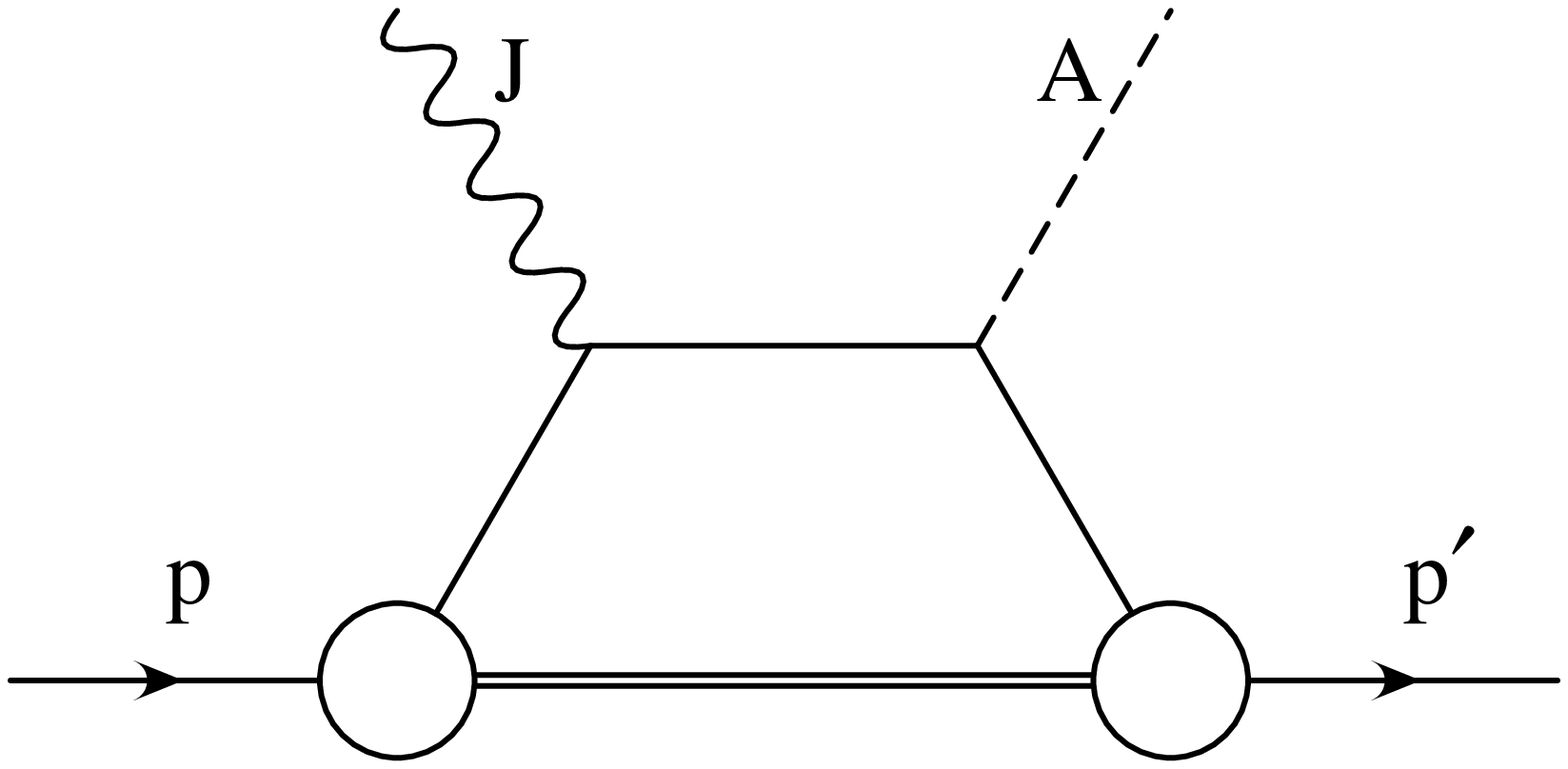,height=3.6in}}
\vskip6pt
\centerline{\psfig{file=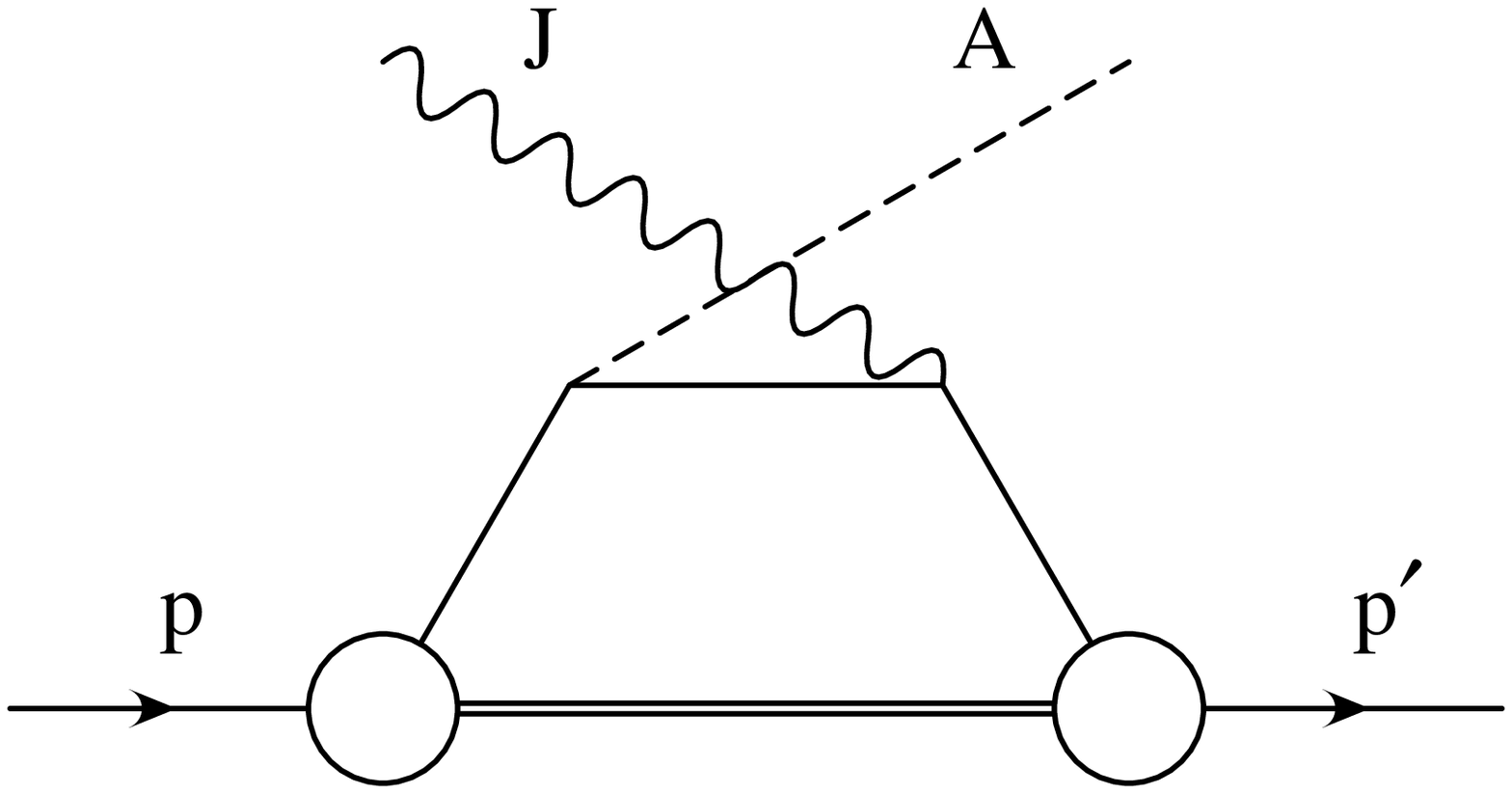,height=3.6in}}
\vskip4pt\centerline{Fig. 3}
\hangindent\parindent{The leading order parton model contributions to $C_{\mu
\nu}$: (a) the direct and (b) the crossed contributions.}}

\vskip3ptplus1fil\vbox{\centerline{\psfig{file=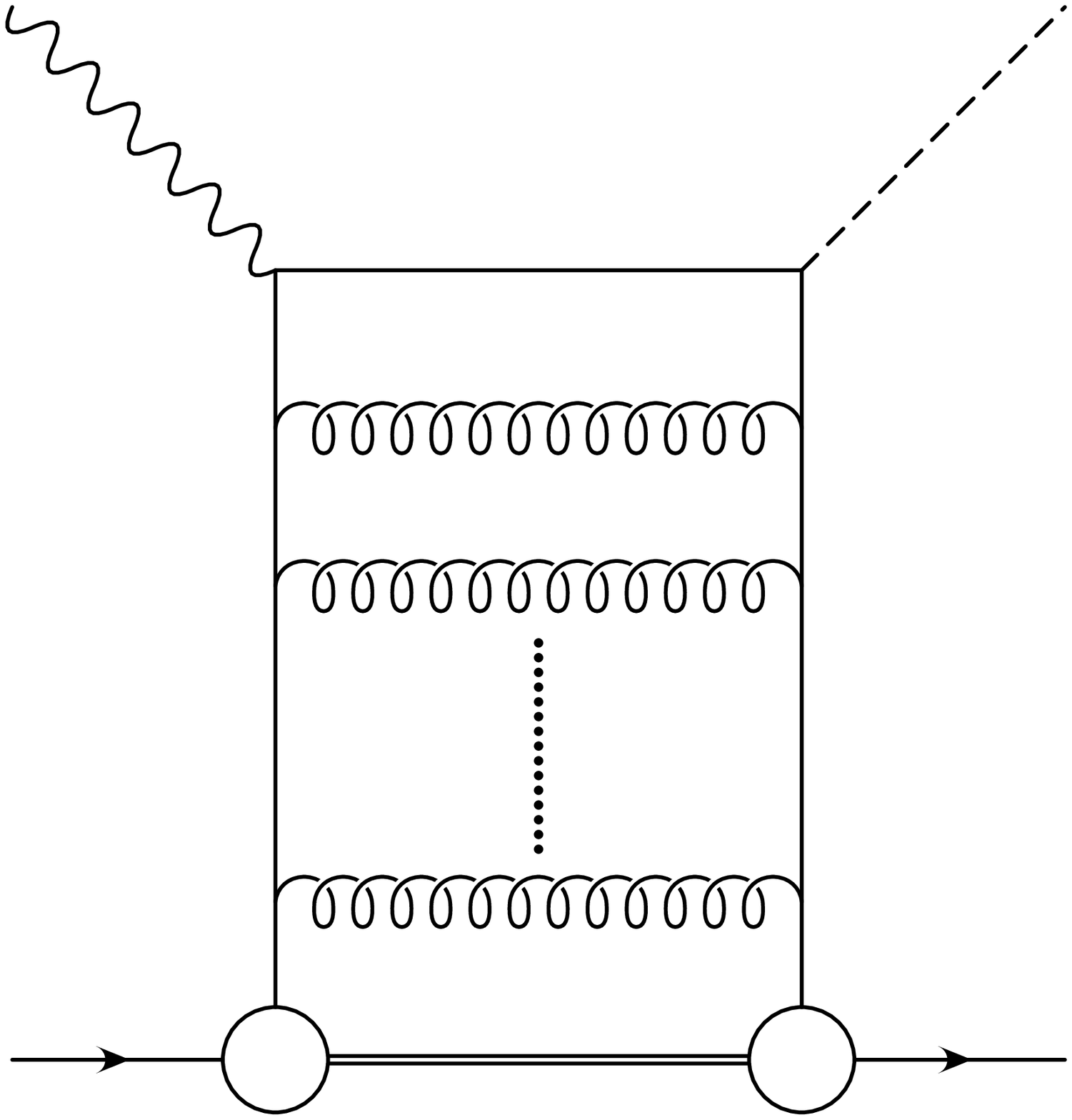,height=7.5in}}
\vskip4pt\centerline{Fig. 4(a)}
\hangindent\parindent{Ladder graph contributions which lead to the leading
logarithmic corrections to the parton model.}}

\vskip3ptplus1fil\vbox{\centerline{\psfig{file=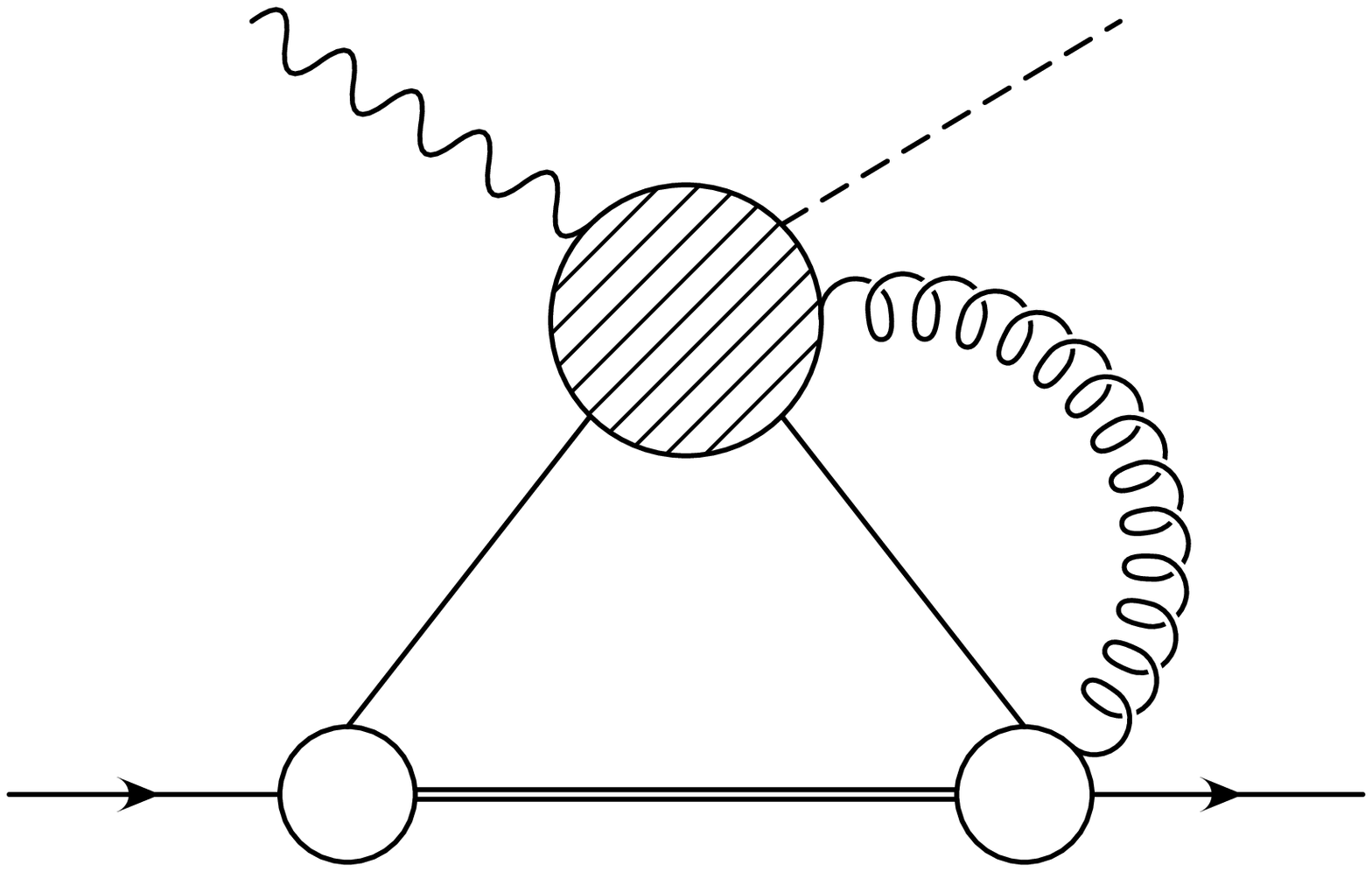,rheight=3.0in}}
\vskip50pt\centerline{Fig. 4(b)}
\hangindent\parindent\centerline{Typical gluon correction that has no leading
logarithmic correction.}}

\vskip3ptplus1fil\vbox{\centerline{\psfig{file=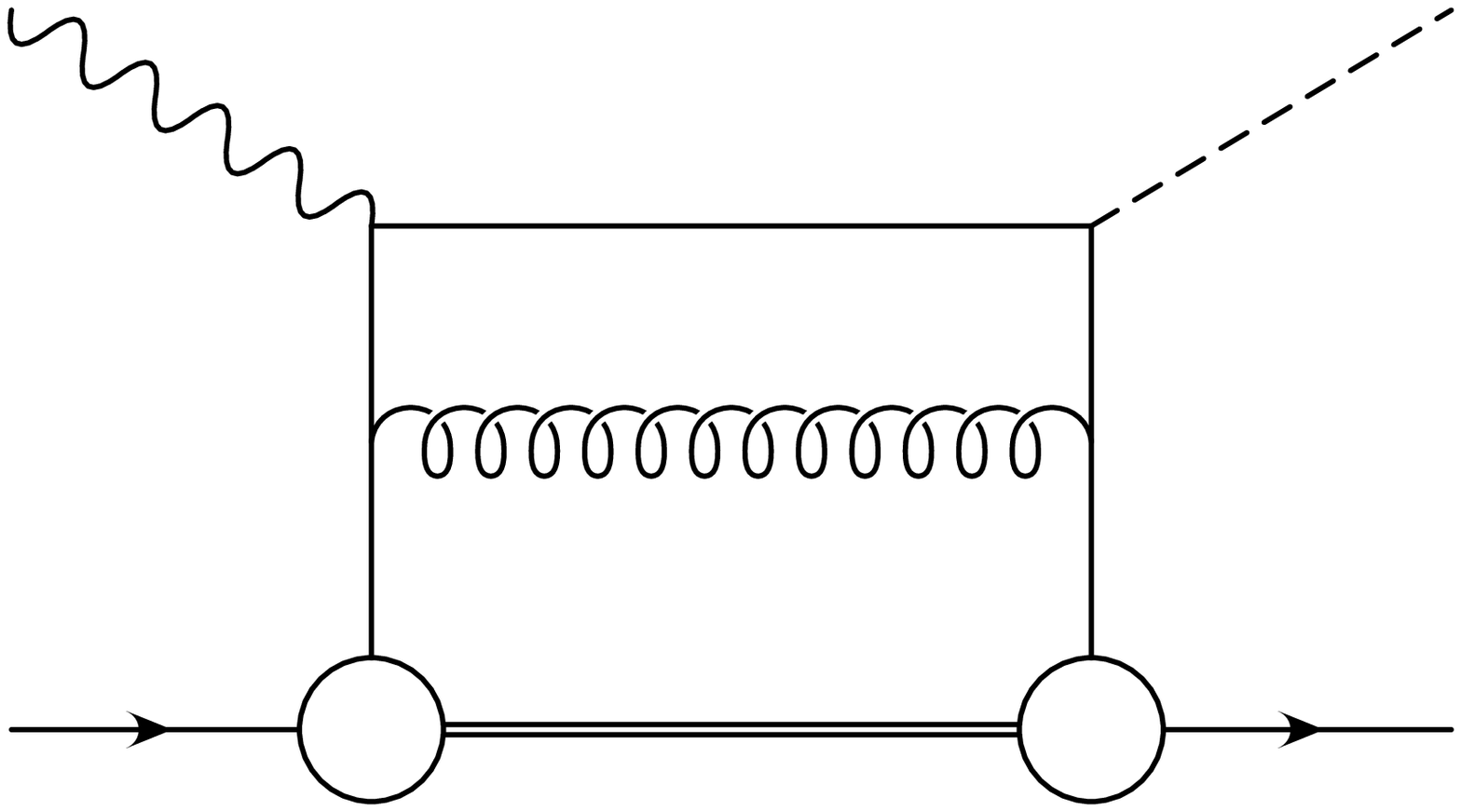,rheight=3.0in}}
\vskip40pt\centerline{Fig. 5}
\hangindent\parindent\centerline{Lowest order leading logarithm gluon
correction.}}

\vskip3ptplus1fil\vbox{
   \centerline{\psfig{file=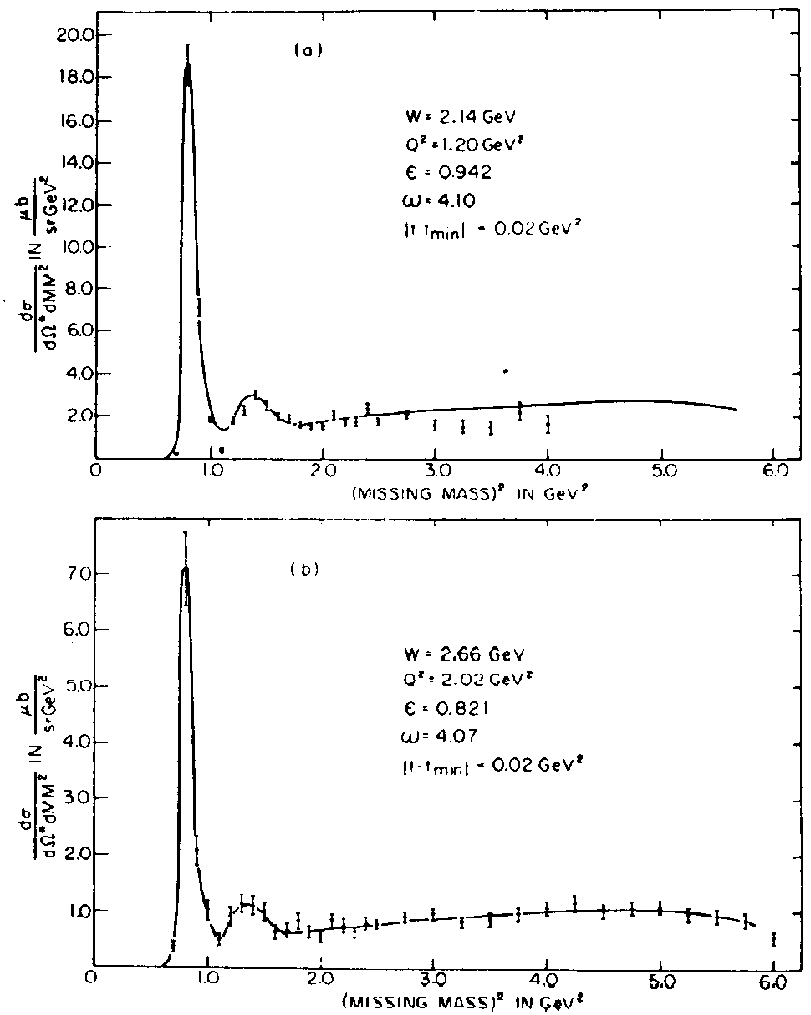,height=4.5in}}
   \vskip1.0in
   \centerline{Fig. 6}
   \hangindent\parindent
   {The virtual photoproduction cross-section at two different values of
$W=\sqrt{s}$ but at the same value of $\omega\equiv 1/x$; the data are taken
from \rlnine .  According to eqs. (4.1) and (4.2) these should be identical
except for a scale factor $W^{4}$.}}

\vskip3ptplus1fil\vbox{\centerline{\psfig{file=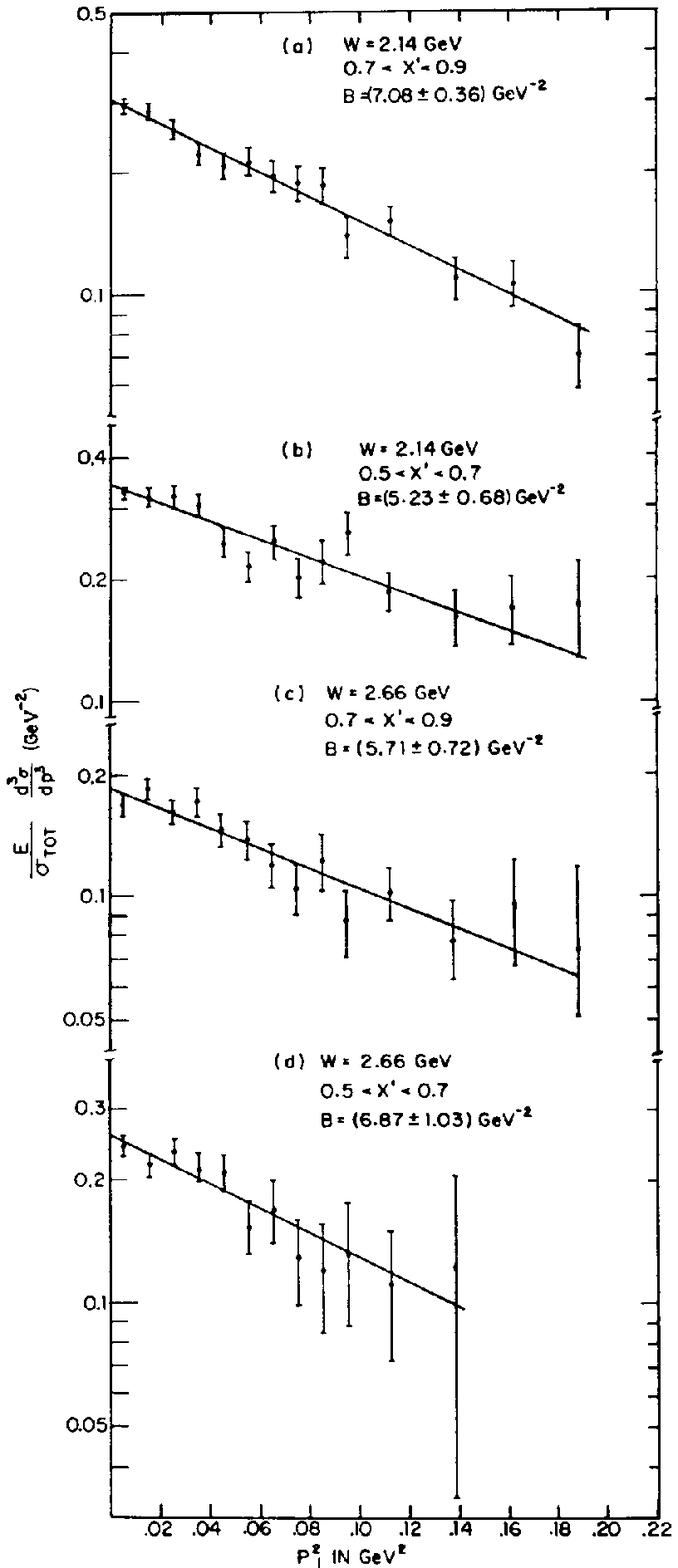,height=6.5in}}
\vskip1.0in\centerline{Fig. 7}
\hangindent\parindent{Transverse-momentum distribution of the produced pions
for two regions of longitudinal pion momentum.  The similarity of the data at
different values of $W$ are in agreement with the scaling argument.}}

\listrefs
\end